\documentclass[a4paper,11pt]{article}
\usepackage{pos}

\usepackage{amsthm}
\usepackage{amsmath}
\usepackage{amssymb}

\title{Phenomenology of scotogenic-like 3-loop neutrino mass models}

\author[a]{Asmaa~Abada}
\author[b]{Nicol\'{a}s Bernal}
\author[c,d,e]{Antonio E. C\'{a}rcamo Hern\'{a}ndez}
\author[e,f]{Sergey Kovalenko}
\author*[e,i]{T\'{e}ssio B. de Melo}
\author[g,h]{Takashi Toma}

\affiliation[a]{P\^ole Th\'eorie, Laboratoire de Physique des 2 Infinis Ir\`ene Joliot Curie (UMR 9012)\\
CNRS/IN2P3, 15 Rue Georges Clemenceau, 91400 Orsay, France\\}

\affiliation[b]{New York University Abu Dhabi, PO Box 129188, Saadiyat Island, Abu Dhabi, United Arab Emirates\\}

\affiliation[c]{Universidad T\'{e}cnica Federico Santa Mar\'{\i}a, Casilla 110-V, Valpara\'{\i}so, Chile\\}

\affiliation[d]{Centro Cient\'{\i}fico-Tecnol\'{o}gico de Valpara\'{\i}so, Casilla 110-V, Valpara\'{\i}so, Chile\\}

\affiliation[e]{Millennium Institute for Subatomic Physics at the High Energy Frontier (SAPHIR),\\
Fern\'andez Concha 700, Santiago, Chile\\}

\affiliation[f]{Center for Theoretical and Experimental Particle Physics - CTEPP,\\
Facultad de Ciencias Exactas, Universidad Andres Bello, Fernandez Concha 700, Santiago, Chile\\}

\affiliation[g]{Institute of Liberal Arts and Science, Kanazawa University, Kanazawa 920-1192, Japan\\}

\affiliation[h]{Institute for Theoretical Physics, Kanazawa University, Kanazawa 920-1192, Japan\\}

\affiliation[i]{Universidad Vi\~{n}a del Mar, Escuela de Ciencias, Agua Santa 7055, Rodelillo, Vi\~{n}a del Mar, Chile\\}

\emailAdd{asmaa.abada@ijclab.in2p3.fr}
\emailAdd{nicolas.bernal@nyu.edu}
\emailAdd{antonio.carcamo@usm.cl}
\emailAdd{sergey.kovalenko@unab.cl}
\emailAdd{tessio.melo@uvm.cl}
\emailAdd{toma@staff.kanazawa-u.ac.jp}

\abstract{
In this talk, we discuss the phenomenology of radiative 3-loop seesaw models. The 3-loop suppression allows the new particles to have masses at the TeV scale, along with relatively large Yukawa couplings, while retaining consistency with neutrino masses and mixing, as observed in neutrino oscillation experiments. This leads to a rich phenomenology, especially in searches for charged lepton flavor violation, where the models predict sizable rates, well within future experimental reach. The models provide viable fermionic or scalar dark matter candidates, as is typical within the scotogenic paradigm. We discuss specific realizations in which the W-mass anomaly and the baryon asymmetry of the Universe can be accommodated, while complying with current constraints imposed by electroweak precision observables, charged-lepton flavor violation and neutrinoless double-beta decay.
}

\FullConference{42nd International Conference on High Energy Physics (ICHEP2024)\\
18-24 July 2024\\
Prague, Czech Republic\\}

\begin{document}
\maketitle

\section{Introduction}

Scotogenic neutrino mass models are testable extensions of the SM which explain the tiny neutrino masses, with seesaw mediators playing an important role in successfully accommodating the observed amount of Dark Matter 
(DM)~\cite{Cai:2017jrq,Arbelaez:2022ejo,Baek:2017qos}. In radiative seesaw models where the neutrino masses are generated at the 1-loop level, either very suppressed Yukawa couplings or unnaturally small mass splittings between the CP-even and CP-odd neutral scalar mediators are required. 
In this work, we investigate models where light active-neutrino masses are generated at the 3-loop level, offering a more natural explanation for their smallness. The first model, described in Section~\ref{Sec:model_1}, is a direct extension of the scotogenic model~\cite{Tao:1996vb,Ma:2006km,Abada:2022dvm,deMelo:2023yqg}, while the second, covered in Section~\ref{Sec:model_2}, incorporates an inverse seesaw (ISS) structure~\cite{Mohapatra:1986bd,Abada:2023zbb,Abada:2024xpf}. After detailing the field content and symmetries, we explore their key phenomenological implications, including charged-lepton flavor violation (cLFV) and electroweak precision observables. Furthermore, we demonstrate that in addition to explaining neutrino masses and dark matter, the first model can address the W-mass anomaly, while the second can account for the baryon asymmetry of the Universe (BAU) through leptogenesis. Our conclusions are summarized in Section~\ref{Sec:conclusions}.

\section{Model 1}
\label{Sec:model_1}

We begin by discussing a 3-loop neutrino mass model that is more akin to the original scotogenic model~\cite{Tao:1996vb,Ma:2006km}. Its particle content features two RH neutrinos $N_{R_k}$, an inert scalar doublet $\eta$, and four scalar singlets $\sigma$, $\rho$, $\varphi$, $\zeta$. It has an extended symmetry group, including a spontaneously broken global symmetry $U(1)'$ and a preserved discrete symmetry $\mathbb{Z}_2$. The new particles are all odd under the $\mathbb{Z}_2$ symmetry, except for $\sigma$, which breaks the $U(1)'$ symmetry at the TeV scale. The lightest $\mathbb{Z}_2$-odd state among the electrically neutral scalars and the two states $N_{R_{k}}$ is thus a viable DM candidate.

\begin{figure}[b]
    \begin{center}
        \includegraphics[width=0.45\textwidth]{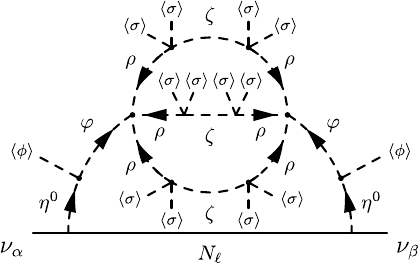} 
    \end{center}
    \caption{Diagram for neutrino masses, with $\ell=1$, 2 and $\protect\alpha,\, \protect\beta = e,\, \protect\mu,\, \protect\tau$.}
    \label{Neutrinodiagram}
\end{figure}

The new interactions, consistent with the symmetries and particle content, are given by the following Lagrangian:
\begin{equation*}
\begin{split}
\label{eq:lagrangian2}
- \mathcal{L} \supset & y_{u \phi}^{ij}\, \bar{q}_{iL} \widetilde{\phi} u_{jR} + y_{d \phi}^{ij}\, \bar{q}_{iL} \phi d_{jR} + y_{l \phi}^{ij}\, \bar{\ell}_{iL} \phi \ell_{R_j} + y_\eta^{ik}\, \bar{\ell}_{iL} \widetilde{\eta} N_{R_{k}} + M_{N_{R}}^{kr}\, \bar{N}_{R_{k}} N_{R_{r}}^C + \mathrm{H.c.} \\
& + \lambda _{15}\left( \rho \zeta \sigma ^{2}+\mathrm{H.c.}\right) + \lambda_{14}\left( \varphi \rho ^{3}+\mathrm{H.c.}\right) + A\left[ (\eta ^{\dagger }\phi )\varphi +\mathrm{H.c.}\right] .
\end{split}
\end{equation*}
These interactions allow for a 3-loop contribution to the neutrino masses, as depicted in Fig. \ref{Neutrinodiagram}. As the imposed symmetries forbid a Dirac mass term at tree level, and also preclude 1-loop and 2-loop-level contributions, the 3-loop diagram provides the leading contribution to neutrino masses in this model.

\paragraph{$W$ boson mass}

\begin{figure}[t]
    \centering
    \includegraphics[width=0.445\linewidth]{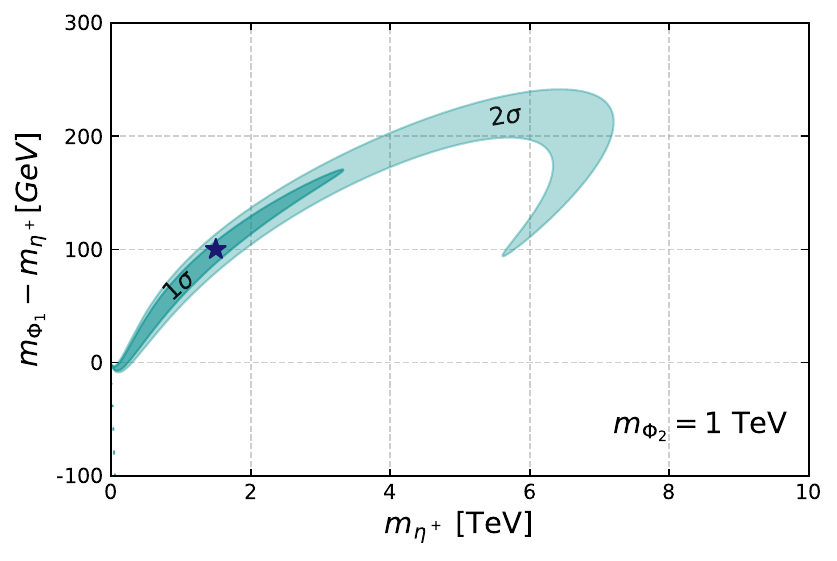}
    \includegraphics[width=0.525\linewidth]{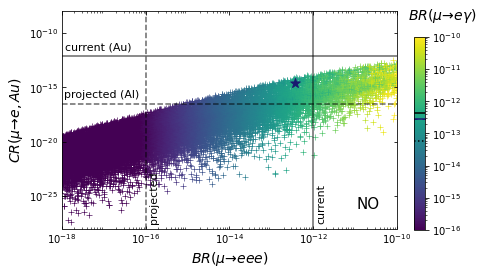}
    \caption{\textit{Left panel}: regions in the $m_{\Phi_1} - m_{\protect\eta^+}$ versus $m_{\protect\eta^+}$ parameter space that accommodate the CDF measurement of the $W$ mass at $1$-$\protect\sigma$ and $2$-$\protect\sigma$. \textit{Right panel}: correlation among rates of cLFV processes. 
    The current upper bounds are indicated by the black full lines, while the future sensitivities, by the black dashed lines.}
    \label{stuCDFplot}
\end{figure}

New physics contributions to the SM oblique corrections, usually expressed in terms of the $S$, $T$, $U$ parameters, shift the $W$ gauge boson mass according to: 
\begin{equation}
M _W ^2 = \left( M _W ^2 \right) _\text{SM} + \frac{\alpha _\text{EM} \left( M _Z \right) \cos ^2 \theta _W \, M _Z ^2}{\cos ^2 \theta _W - \sin ^2 \theta _W} \left[ - \frac{S}{2} + \cos ^2 \theta _W \, T + \frac{\cos ^2 \theta _W - \sin ^2 \theta _W}{4\, \sin ^2 \theta _W}\, U \right] .
\end{equation}
In this model, the oblique corrections are affected by the presence of the extra scalars from the inert doublet, thus providing the ingredients to explain the measured value of the $W$ mass by the CDF collaboration~\cite{CDF:2022hxs}. As shown in the left panel of Fig.~\ref{stuCDFplot}, the CDF anomaly can be accomodated with scalar masses in the TeV scale, as long as the mass splitting among the charged and neutral scalars is within a few hundred GeV.

\paragraph{Charged lepton flavor violation}

This model can be tested in experiments that search for cLFV processes, which arise at 1-loop level from the exchange of charged scalars $\eta ^\pm$ and RH neutrinos $N_{R_k}$, including the radiative decays $\mu \rightarrow e\gamma $ and $\mu \to e e e $, and $\mu-e$ conversion in atomic nuclei. We scan the parameter space of the model computing the corresponding cLFV rates. The result is shown in the right panel of Fig.~\ref{stuCDFplot}. While current cLFV bounds provide stringent constraints to the model, there is a large portion of the parameter space that pass the bounds and is within the reach of future $\mu-e$ conversion and $\mu \to e e e$ experiments. A representative point of this parameter space is the benchmark point shown in Fig.~\ref{stuCDFplot}, marked as a blue star, which besides complying with all the current constraints, account for the $W$ mass anomaly, is within the reach of future cLFV experiments and accomodates neutrino masses and DM relic abundance.


\section{Model 2}
\label{Sec:model_2}

In this section, we describe an instance of a 3-loop neutrino mass model which features an inverse seesaw structure. The SM gauge symmetry is extended with the global symmetry $U(1)' \otimes \mathbb{Z}_2$, whereas the extra particle content consists of three scalars fields $\varphi_{1,2}, \sigma$, two vector-like neutral leptons $\Psi_{1,2}$ and two left-handed Majorana neutrinos $\Omega_{1,2}$. 
The global $U(1)'$ symmetry is spontaneously broken at the TeV scale by the VEV of $\langle\sigma\rangle$ down to a residual preserved $\mathbb{Z}_4$ symmetry, resulting in particle masses at the TeV scale. The Yukawa interactions relevant to the neutrino mass generation are given by 
\begin{align} \label{Eq:Lag3}
- & \mathcal{L} _Y = \left(y_{\nu}\right)_{ik}\, \overline{l}_{iL}\, \widetilde{\phi}\, \nu_{kR} + M_{nk}\, \overline{\nu}_{nR}\, N_{kR}^C + \left(y_{N}\right)_{nk}N_{nR}\, \varphi_1^*\, \overline{\Psi_{kR}^C} \notag \\
&+ \left(y_{\Omega}\right)_{nk}\, \Psi_{nL}^C\, \varphi_2\overline{\Omega}_{kL} + \left(y_{\Psi}\right)_{nk}\, \overline{\Psi}_{nL}\, \sigma\, \Psi_{kR} + \left(m_{\Omega}\right)_{nk}\, \overline{\Omega}_{kL}\, \Omega_{nL}^C + \text{H.c.}
\end{align}

\begin{figure}[!t]
    \begin{center}
        \includegraphics[width=0.45\textwidth]{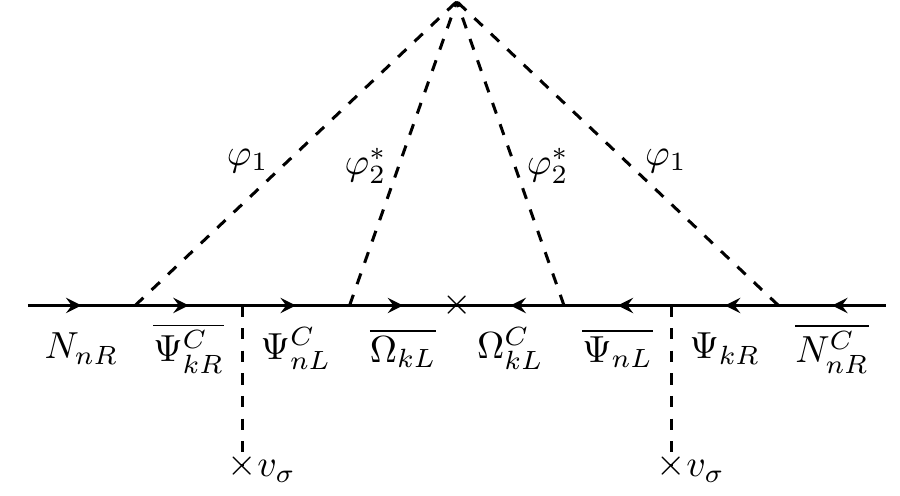} 
    \end{center}
    \caption{Diagram for the lepton number violating Majorana mass in the ISS, with $n, k = 1, 2$.}
    \label{diagram_inv}
\end{figure}

A lepton number violating Majorana mass term $\mu_{nk}\, \overline{N_{nR}}\, N_{kR}^C$ is induced at the three-loop level according to the diagram shown in Fig.~\ref{diagram_inv}. The physical neutrino spectrum is composed of three light active neutrinos and four neutral states, which form two pairs of pseudo-Dirac neutrinos, denoted as $N ^\pm _k$. The mass splitting of the quasi-degenerate pseudo-Dirac pairs is proportional to the small Majorana mass scale $\mu$.

\paragraph{Dark Matter, Leptogenesis and Charged lepton flavor violation}

For definiteness, we assume $\Psi _{1R}$ is the lightest odd state and, therefore, the DM candidate of the model. For masses in the GeV-to-TeV range and large Yukawa couplings, the DM can be thermally produced in the early Universe.
In the presence of out-of-equilibrium CP and lepton number violating decays of $N_k^{\pm}$, the baryon asymmetry of the Universe (BAU) can also be generated via leptogenesis. Assuming that $| M_{N _1 ^\pm} | \ll | M_{N _2 ^\pm} | $ so that only the first generation of $N_k^{\pm}$ contribute to the BAU, the correct lepton asymmetry parameter can be obtained both in the strong and weak washout regimes thanks to the small sppliting between $N _1 ^+$ and $N _1 ^-$, which is proportional to the 3-loop suppressed Majorana $\mu$ parameter.

\begin{figure}[b]
    \begin{center}
    \includegraphics[width=0.47\linewidth]{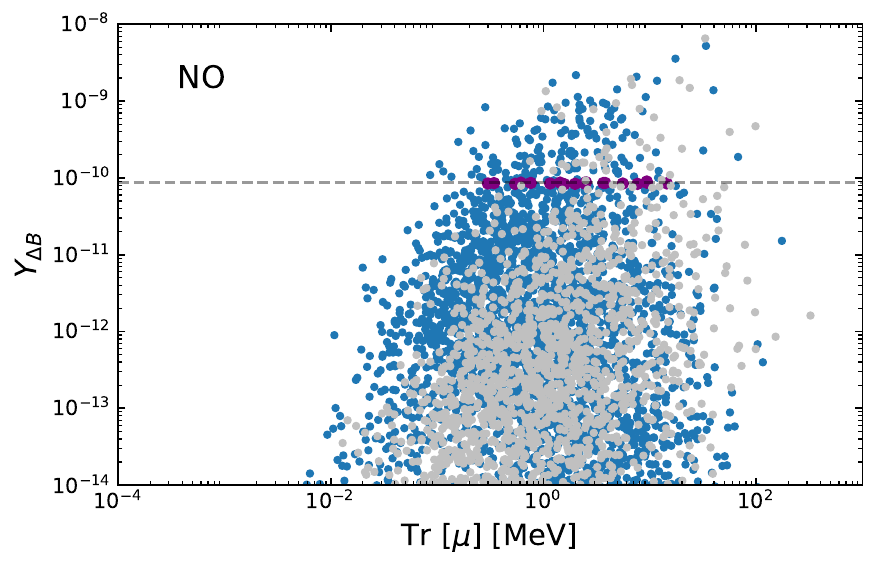}
    \includegraphics[width=0.505\linewidth]{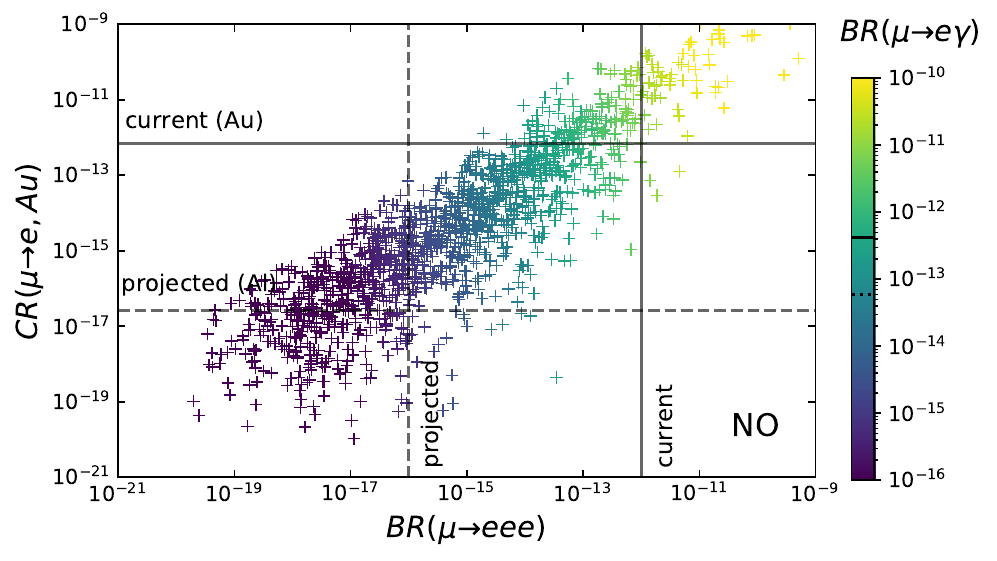}
    \end{center}
    \caption{
    \textit{Left panel}: baryon asymmetry parameter $Y_{\Delta B}$ versus the trace of the Majorana matrix $\mu$. All points comply with the DM relic abundance and neutrino oscillation data assuming normal ordering. The purple points have the correct $Y_{\Delta B}$ within $3 \sigma$, while the gray points are excluded by cLFV. The values of $\mu$ in the keV-MeV range are favored by leptogenesis. \textit{Right panel}: Correlation among cLFV observables. Points comply with the BAU, DM relic abundance and neutrino oscillation data assuming normal ordering. Current bounds are shown as full lines, while projections are shown as dashed lines.}
    \label{yvsmN}
\end{figure}

In the left panel of Fig.~\ref{yvsmN} we plot the baryon asymmetry parameter $Y_{\Delta B}$ against the trace of the $\mu$ Majorana matrix. The purple points in this plot correspond to the values of $Y_{\Delta B}$ within the experimentally allowed range at $3 \sigma$~CL, while the gray points represent excluded points by cLFV constraints. 
In the right panel of 
Fig.~\ref{yvsmN} we show the correlation among the cLFV rates and the current and future constraints, showing that this model also predicts measurable signals in future cLFV experiments. In this plot, all points reproduce the correct baryon asymmetry parameter and DM relic abundance.

\section{Conclusion} \label{Sec:conclusions}


We have investigated two models where active neutrino masses are radiatively generated at the three-loop level, providing a natural explanation for their smallness. The new particle masses can remain at the TeV scale without requiring fine-tuning of Yukawa couplings. Both models are consistent with experimental constraints, including those from neutrino oscillation data, neutrinoless double-beta decay, dark matter relic abundance, charged lepton flavor violation, and electron-muon conversion processes. The first setup, an extended version of the original scotogenic model, offers a simple solution to the W-mass anomaly, while the second model, which features an inverse seesaw, enables low-scale resonant leptogenesis, providing a mechanism to generate the baryon asymmetry of the Universe. Furthermore, we found that both models predict significant rates for charged lepton flavor-violating processes, such as $\mu\to e\gamma$ and $\mu\to eee$, as well as for electron-muon conversion. These rates fall within the reach of upcoming experimental sensitivities, making these models testable in the near future.

\section*{Acknowledgments}

\noindent
The results here presented are based on Refs.~\cite{Abada:2022dvm,Abada:2023zbb}. 
NB received funding from the Spanish FEDER/MCIU-AEI under grant FPA2017-84543-P. 
This project has received funding and support from the European Union's Horizon 2020 research and innovation programme under the Marie Sk{\l}odowska-Curie grant agreement
No.~860881 (H2020-MSCA-ITN-2019 HIDDeN) and from the Marie Sk{\l}odowska-Curie Staff Exchange grant agreement No 101086085 ``ASYMMETRY''. A.E.C.H.. and S.K. are supported
by ANID-Chile FONDECYT 1170171, 1210378, 1230160, ANID PIA/APOYO AFB230003, and Proyecto Milenio- ANID: ICN2019\_044. TBM acknowledges ANID-Chile grant FONDECYT No. 3220454 for fnancial support.
This work was supported by the JSPS Grant-in-Aid for Scientific Research KAKENHI Grant No. JP20K22349 (TT).

%

\bibliographystyle{JHEP}
\bibliography{biblio}

\end{document}